\documentclass[conference]{IEEEtran}
\IEEEoverridecommandlockouts
\usepackage{cite}
\usepackage{amsmath,amssymb,amsfonts}
\usepackage{algorithm}
\usepackage{algorithmic}
\usepackage{graphicx}
\usepackage{textcomp}
\usepackage{xcolor}
\usepackage{multirow}
\usepackage{url}
\usepackage{flushend}
\def\BibTeX{{\rm B\kern-.05em{\sc i\kern-.025em b}\kern-.08em
    T\kern-.1667em\lower.7ex\hbox{E}\kern-.125emX}}
\begin{document}

\title{A Game Between the Defender and the Attacker for Trigger-based Black-box Model Watermarking}

\author{Chaoyue Huang and Hanzhou Wu\\
Shanghai University, Shanghai 200444, China\\
\{buttercup123, hanzhou\}@shu.edu.cn}

\maketitle

\begin{abstract}
Watermarking deep neural network (DNN) models has attracted a great deal of attention and interest in recent years because of the increasing demand to protect the intellectual property of DNN models. Many practical algorithms have been proposed by covertly embedding a secret watermark into a given DNN model through either parametric/structural modulation or backdooring against intellectual property infringement from the attacker while preserving the model performance on the original task. Despite the performance of these approaches, the lack of basic research restricts the algorithmic design to either a trial-based method or a data-driven technique. This has motivated the authors in this paper to introduce a game between the model attacker and the model defender for trigger-based black-box model watermarking. For each of the two players, we construct the payoff function and determine the optimal response, which enriches the theoretical foundation of model watermarking and may inspire us to develop novel schemes in the future.
\end{abstract}

\begin{IEEEkeywords}
DNN watermarking, game theory, security, deep learning, information hiding.
\end{IEEEkeywords}

\section{Introduction}
Digital watermarking hides a secret signal (typically called \emph{watermark}) within another noise-tolerant signal such as image, video, and text. By extracting the watermark from the target watermarked signal probably attacked, we are capable of identifying the ownership of the watermarked signal, which promotes digital watermarking to play a quite important role in protecting the intellectual property (IP) of digital commercial products. In the past three decades, many digital watermarking algorithms have been developed to protect the IP of multimedia content. They often model multimedia to be watermarked as a static signal which can be treated as a sequence of real-valued numbers (also called \emph{cover sequence}). Since these cover elements are highly correlated to each other, individual cover elements can be accurately predicted from the local context. As a result, these signals are easy to model, enabling the watermark to be inserted into the selected components of the signals without reducing their commercial value.

Thanks to the fast development of computer hardware and big data technology in the past decade, deep learning (DL) has achieved great success in many application areas. Especially, as the representative architecture of DL, deep neural networks (DNNs), have brought profound changes to our society due to its superior performance in many vision and natural language related tasks. It is straightforward to apply advanced DNNs to conventional or novel multimedia watermarking systems to further improve their performance, which has been extensively studied in recent years. However, in addition to utilizing DNNs for watermarking signals, a DNN itself should be protected as well since creating a state-of-the-art DNN model requires large-scale well-labeled data, expert knowledge and powerful computational resources. Furthermore, the openness of DNN models enables attackers to easily tamper and sell DNN models. It is very urgent to prevent DNN models from intellectual property infringement, which creates a new research direction called \emph{DNN watermarking} or \emph{model watermarking} \cite{Wu2023CJIG}.

Mainstream DNN watermarking methods can be categorized to \emph{white-box} and \emph{black-box}, depending on the access authority of the watermark extractor. The former enables the watermark extractor to access the internal information of the target model including network structure and parameters. Along this line, a lot of white-box methods are proposed to embed a watermark into the DNN model via modulating the network weights \cite{Uchida2017, WangSPIE2020}, structure \cite{Zhao2021WIFS} or statistics of the internal feature maps \cite{Fan2019}. However, it is often the case in applications that the watermark extractor does not know the internal details of the target model, which limits the usage and motivates scholars to study black-box watermarking. Black-box watermarking assumes that the watermark extractor does not know the internal details, but can interact with the target model. The watermark is extracted by querying the target model with a certain number of carefully crafted trigger samples \cite{adi:2018, Zhao:ISDFS, Wang:Sym2022, Lin2022}. As a special case of black-box, \emph{box-free} watermarking has been studied recently \cite{wuTCSVT2021}. In the case, the watermark extractor does not know the internal details of the target model and cannot interact with the target model. Instead, it uses the model output to carry a watermark for ownership verification, e.g., by retrieving the watermark from any image generated by the target model, one can identify the ownership of the target model and the generated image.

Compared with white-box, black-box watermarking is more common in applications. However, most existing watermarking methods focus on practical design. Few works study the theoretical basis of black-box model watermarking. As a result, the lack of basic research restricts the algorithmic design to a trial-based method or a data-driven technique. In order to tackle with this gap, in this article, we introduce a game between the model defender and the model attacker for black-box classification model watermarking, which typically uses trigger samples for model verification. By designing the payoff function for each of the two players, we determine the optimal strategies for the two players, which may enrich the theoretical foundation of model watermarking and inspire us to develop novel practical methods in the future.

The rest structure of this paper is organized as follows. In Section II, we briefly introduce the necessary preliminaries. Then, in Section III, we present a game between the model defender and attacker for trigger-based black-box model watermarking, for which we determine the optimal strategies for both players. Finally, we conclude this paper in Section IV. 

\section{Preliminaries}
\subsection{Black-box Model Watermarking}
Before game analysis, we firstly describe the general framework of black-box classification model watermarking in this subsection. We here limit the original task of the host model to image classification. That is, the host model can be trained on a normal image dataset $D = \{(\textbf{x}_i, y_i) | 1\leq i\leq |D|\}$ to generate a non-marked model $\mathcal{M}$. Here, $\textbf{x}_i$ is a clean image sample, and $y_i$ represents the ground-truth label of $\textbf{x}_i$. In order to watermark $\mathcal{M}$ under the black-box scenario, a set of trigger samples defined as $D^\text{T} = \{(\textbf{x}_i^\text{T}, y_i^\text{T}) | 1\leq i\leq |D^\text{T}|\}$ should be constructed in advance. Then, $\mathcal{M}$ can be trained from scratch with $D$ and $D^\text{T}$ such that a watermarked model, expressed as $\mathcal{M}_\text{w}$, is generated. It is required that the generalization ability of $\mathcal{M}$ and $\mathcal{M}_\text{w}$ should be close to each other, which can be quantified from a statistical perspective, i.e., 
\begin{equation}
1-\frac{1}{|E|}\sum_{i=1}^{|E|}\delta(\mathcal{M}(\textbf{x}_i^\text{E}),\mathcal{M}_\text{w}(\textbf{x}_i^\text{E})) \leq \Delta,
\end{equation}
where $E = \{(\textbf{x}_i^\text{E}, y_i^\text{E}) | 1\leq i\leq |E|\}$ is the test set, $\delta(x,y) = 1$ if $x = y$ otherwise $\delta(x,y) = 0$, and $\Delta$ is a small positive real number approaching zero. Meanwhile, it is necessary that $\mathcal{M}_\text{w}$ results in a high prediction accuracy on a trigger set $E^\text{T} = \{(\textbf{x}_i^\text{ET}, y_i^\text{ET}) | 1\leq i\leq |E^\text{T}|\}$ so that the ownership of the model can be verified when disputes arise, i.e., 
\begin{equation}
1-\frac{1}{|E^\text{T}|}\sum_{i=1}^{|E^\text{T}|}\delta(\mathcal{M}_\text{w}(\textbf{x}_i^\text{ET}), y_i^\text{ET}) \leq \Delta.
\end{equation}

In this way, black-box model watermarking can be realized.

\subsection{Game Theory}
Game theory is the study of mathematical models of strategic interactions among two or more rational players \cite{game:book}. It has been used in many application fields such as finance, politics and information science. Mathematically, a game consists of $n\geq 2$ players. Each player $i$ has a set of possible strategies, expressed as $S_i$. A pure strategy vector can be then expressed as $\textbf{s} = (s_1, s_2, ..., s_n)$, where $s_i\in S_i$ is the strategy selected by the player $i$. The pure strategy vector $\textbf{s}\subset \times_iS_i$ determines the payoff for each player and the payoff will be different for different players generally. The payoff function for player $i$ can be defined as $u_i(\textbf{s}) = u_i(s_1, s_2, ..., s_n)\in \mathbb{R}$. An \emph{equilibrium} is a strategy vector $\textbf{s}^* = (s_1^*, s_2^*, ..., s_n^*)$ that for each player $1\leq i\leq n$, the strategy $s_i^*\in S_i$ satisfies that
\begin{equation}
\begin{split}
u_i(s_1^*, s_2^*, &..., s_{i-1}^*, s_i^*, s_{i+1}^*, ..., s_n^*) \geq\\
&u_i(s_1^*, s_2^*, ..., s_{i-1}^*, s_i, s_{i+1}^*, ..., s_n^*), \forall s_i\in S_i.
\end{split}
\end{equation}

For the equilibrium, no player can do better by unilaterally changing his strategy. A game may have more than one equilibrium. However, in many games, there may be no equilibrium under condition of using pure strategy only. Fortunately, if mixed strategies where a player chooses probabilities of using various pure strategies are allowed, every game with a finite number of players in which each player can choose from finitely many pure strategies has at least one equilibrium (called \emph{Nash equilibrium}). The equilibrium might be a pure strategy for each player or might be a probability distribution over strategies for each player. The game analyzed here is simultaneous (i.e., static) and requires that each player masters perfect information about the strategy space of others.

\subsection{Related Works}
Existing studies applying game theory to digital watermarking are focused on multimedia signals. They mainly investigate the trade-off between robustness, capacity and system security. For example, Wu \emph{et al.} introduce a two-encoder game related to rate-distortion optimization of reversible watermarking \cite{wuTR2021}. They investigate non-cooperative and cooperative games from the parameterized perspective, and determine the equilibrium strategies for both encoders under constraints. Giboulot \emph{et al.} \cite{EvaTIFS2023} investigate the game played between the steganographer and the steganalyst in a real-world environment, where various factors should be considered. They introduce a two-player non-zero-sum game constrained by the environment composed of multiple cover sources, which can be reduced to a zero-sum game allowing the Nash equilibrium to be determined. More games can be found from references \cite{vaiIJMLC2019, TsaiESWA2011}. These games cannot be directly extended to DNNs as watermarks embedded in DNNs exhibit characteristics that are significantly different from that embedded in multimedia signals.

Recently, game-theoretic approaches have shifted their focus towards adversarial machine learning. In \cite{SamGameSec2021, MerTIFS2024}, the authors introduce a classification game that captures all relevant properties in the competition between the adversary and defender in adversarial machine learning. The game can be instantiated with various defender and adversary strategies, demonstrating good applicability. In \cite{KalEURASIP2024}, the authors present a game between an attacker and a defender in the context of DNN backdoor. Through analysis, the work highlights its potential to enhance DNN security against backdooring. Recalling the fact that black-box DNN watermarking can be achieved using backdooring, it is straightforward to think of modeling DNN watermarking as a game in the context of DNN backdoor. To this end, we investigate trigger-based model watermarking and present a partial cooperation game framework in this paper.

Unlike previous approaches mainly focusing on competitive dynamics, this model examines scenarios where the attacker and the defender engage in cooperation on maintaining model performance. Our model expands the analytical horizon in adversarial machine learning by integrating cooperative elements into the adversarial context, thus offering novel insights for the design of model watermarking strategies. We are committed to bridge the gap between theoretical advancements and practical implementations, thereby contributing to the development of more secure and resilient machine learning systems.

\section{Game Analysis}
\subsection{Game Formulation}
Let $\mathcal{M}_0$ be the original model trained with normal samples, i.e., $\mathcal{M}_0$ is \emph{non-marked}. Let $\mathcal{M}_1$, $\mathcal{M}_2$, ..., $\mathcal{M}_N$ be $N$ marked versions of $\mathcal{M}_0$ by applying different watermark embedding strategies which are characterized by the proportion of trigger samples used for model training equivalent to watermark embedding $\alpha_1, \alpha_2, ..., \alpha_N$. For example, $\alpha_0 = 0$ is corresponding to $\mathcal{M}_0$ indicating that no trigger samples were used for training the model. For each $\mathcal{M}_i$, it can be characterized by $p_i\in [0, 1]$ and $q_i\in [0, 1]$, where $p_i$ is the classification accuracy of $\mathcal{M}_i$ on the normal test set and $q_i$ is the detection accuracy of $\mathcal{M}_i$ on the trigger test set. For $p_i$ and $q_i$, they are expected to be as high as possible generally. However, the performance of the model will change depending on the attacker's attack strategy.  

Mathematically, the strategy space of the defender \emph{Alice} can be expressed as $\mathcal{S}_A = \{\alpha_0, \alpha_1, ..., \alpha_N\}$, and the strategy space of the model attacker \emph{Bob} is given by $\mathcal{S}_B = \{\beta_0, \beta_1, ..., \beta_M\}$, where $\beta_j$ means the intensity of
 the $j$-th attack, which quantifies the degree to which \emph{Bob} attacks the model. It is assumed that $0 = \alpha_0 < ... < \alpha_N \leq 1$ and $0 = \beta_0 < ... < \beta_M \leq 1$, e.g., $\beta_0$ means no attack is applied to the model whereas $\beta_M$ is corresponding to the strongest attack. We define the robustness of the model $\mathcal{M}_i$ to the $j$-th attack as $r_{i,j} \in [0, 1]$ which shows how much the performance of $\mathcal{M}_i$ drops on both the original task and the watermark detection task after an attack. We then define the computing success rate (CSR) of $\mathcal{M}_i$ after applying the $j$-th attack with $\beta_j$ as:
 \begin{equation}
 \text{CSR}_{i,j} = (1-\beta_j)\left[(1-\alpha_i)p_i+\alpha_iq_i\right] + \beta_jr_{i,j},
 \end{equation}
for which the performance of the model is unchanged if \emph{Bob} does not attack the model (i.e., $\beta_j = 0$), and the performance only depends on the robustness of the model to the $j$-th attack if $\beta_j = 1$. The attack success rate (ASR) of the $j$-th attack on $\mathcal{M}_i$ can be therefore defined as:
\begin{equation}
\text{ASR}_{i,j} = 1 - r_{i,j}.
\end{equation}

Inspired by \cite{SamGameSec2021, MerTIFS2024}, we take into account economic costs and benefits for defining the payoff function, i.e.,
\begin{equation}
\text{U}^\text{Player} = - I^\text{Player} - O^\text{Player} + R^\text{Player}, 
\end{equation}
where \text{Player} $\in$ \{Alice, Bob\}, $I^\text{Player}$ represents the initial cost which is mandatory and independent of whether an attack or defense strategy is adopted, $O^\text{Player}$ represents the ongoing cost which is proportional to the intensity of attack or defense, and $O^\text{Player}$ represents the reward which should be proportional to the success degree of employing a defense or attack strategy.

Accordingly, based on Eq. (4, 5, 6), the payoff function for \emph{Alice} is further defined as:
\begin{equation}
\text{U}_{i,j}^\text{Alice} = - I^\text{Alice} - O_{i,j}^\text{Alice} + R^\text{Alice}_{i,j},
\end{equation}
and the payoff function for \emph{Bob} is given by:
\begin{equation}
\text{U}_{i,j}^\text{Bob} = - I^\text{Bob} - O_{i,j}^\text{Bob} + R^\text{Bob}_{i,j}.
\end{equation}

Both $I^\text{Alice} = I^\text{def}$ and $I^\text{Bob} = I^\text{att}$ are constant, independent of the actions taken by \emph{Alice} and \emph{Bob}. Although the ongoing cost should change depending on the watermark embedding or attack strategy, we here simplify it as a linear function of the intensity of watermark embedding or attack. On the one hand, it would facilitate the calculation of the equilibrium. On the other hand, we parameterize the strategy for both players into a single parameter respectively, making it reasonable to model the ongoing cost as a linear function of the parameter. So, we rewrite $O_{i,j}^\text{Alice}$ and $O_{i,j}^\text{Bob}$ as $O_{i,j}^\text{Alice} = \alpha_i O^\text{def}$ and $O_{i,j}^\text{Bob} = \beta_j O^\text{att}$ where both $O^\text{def}$ and $O^\text{att}$ are constant. It can be easily inferred that when $\alpha_i$ and $\beta_j$ increase, the ongoing costs will increase. Both players expect to maintain the performance of the model on the original task. However, they are competing with each other on watermark detection. The performance of the model on its original task should be no less than a threshold for both players generally. Otherwise, the model will lose its commercial value. The degradation degree of the performance of the model $\mathcal{M}_i$ on its original task is affected by $\alpha_i$ and $\beta_j$. For both players, the performance degradation will produce a negative reward. For \emph{Alice}, successfully defending the model results in a positive reward, whereas a failed defense results in a negative reward. For \emph{Bob}, successfully attacking the model results in a positive reward, whereas a failed attack results in a negative reward. Therefore, we rewrite $R^\text{Player}_{i,j}$ as: 
\begin{equation}
R^\text{Player}_{i,j} = - R^\text{Player}_{-} + R^\text{Player}_{+},
\end{equation}
where the first term denotes the negative reward and the second term denotes the positive reward. We quantify the performance degradation of the model after attack on the original task as a linear function $\alpha_i$ and $\beta_j$, indicating that the reward caused by model attack and defense can be simplified as a linear function of $\alpha_i$ and $\beta_j$. Therefore, we define $R^\text{Player}_{i,j}$ as:
\begin{equation}
R^\text{Alice}_{i,j} = - (\text{COO}_{i,j} + \overline{\text{CSR}_{i,j}}) R^\text{def}_{-} + \text{CSR}_{i,j}R^\text{def}_{+}
\end{equation}
and
\begin{equation}
R^\text{Bob}_{i,j} = - (\text{COO}_{i,j} +  \overline{\text{ASR}_{i,j}}) R^\text{att}_{-} + \text{ASR}_{i,j}R^\text{att}_{+},
\end{equation}
where $R^\text{def}_{-}$, $R^\text{def}_{+}$, $R^\text{att}_{-}$, $R^\text{att}_{+}$ are all constant, $\text{COO}_{i,j}$ is a linear combination of $\alpha_i$ and $\beta_j$, $\overline{\text{CSR}_{i,j}} = 1 - \text{CSR}_{i,j}$, and $\overline{\text{ASR}_{i,j}} = 1 - \text{ASR}_{i,j}$.

\subsection{Model Simplification}
Determining the equilibrium points for \emph{Alice} and \emph{Bob} based on Eq. (4) is difficult since the underlying relationship between $p_i$, $q_i$ and $\alpha_i$ is not clear. However, we can find that
\begin{equation}
0\leq \text{min}\{p_i, q_i\}\leq (1-\alpha_i)p_i + \alpha_iq_i \leq \text{max}\{p_i, q_i\} \leq 1.
\end{equation}

Both $p_i$ and $q_i$ should depend on the defense strategy taken by \emph{Alice}, i.e., $\alpha_i$. This indicates that the term $(1-\alpha_i)p_i + \alpha_iq_i$ is equivalent to a function only depending on $\alpha_i$. It allows us to draw out the following equation:
\begin{equation}
(1-\alpha_i)p_i + \alpha_iq_i = 1 - \lambda_i\alpha_i,
\end{equation}
where $\lambda_i$ is a coefficient. Although $\lambda_0$, $\lambda_1$, ..., $\lambda_{N-1}$ and $\lambda_{N}$ can be different from each other, fine-tuning the model further enables us to make the reasonable assumption: 
\begin{equation}
\lambda_0 = \lambda_1 = ... = \lambda_N = \lambda,
\end{equation}
where $\lambda$ is a constant. Hence, Eq. (4) can be rewritten as:
\begin{equation}
\text{CSR}_{i,j} = (1-\beta_j)(1 - \lambda\alpha_i) + \beta_jr_{i,j}.
\end{equation}

Though $\text{COO}_{i,j}$ can be expressed as different combinations of $\alpha_i$ and $\beta_j$, we assign the same weight to them for fair analysis, that is, $\text{COO}_{i,j} = k \alpha_i + k \beta_j$, where $k$ is a predetermined parameter. Taking into account an arbitrary pair of strategies $\alpha_i < \alpha_j$ for \emph{Alice} and $\beta_s < \beta_t$ for \emph{Bob}, from the defensive perspective, $\alpha_j$ is intuitively better than $\alpha_i$ for \emph{Alice} as long as $\alpha_j$ is effective enough. Similarly, $\beta_s$ is intuitively better than $\beta_t$ for \emph{Bob}. This indicates that no matter what strategy the opponent adopts, the present player tends to choose the strongest and more effective defensive strategy or the weakest but effective attack strategy. It means we can further reduce the number of pure strategies for both \emph{Alice} and \emph{Bob} to only $2$ to simplify the game. Along this direction, in the following, we consider that there are only two pure strategies for both players for better analysis.

\subsection{Optimal Responses for the Defender and the Attacker}
We rewrite Eq. (7, 8) as:
\begin{equation}
\begin{split}
\text{U}_{i,j}^\text{Alice} = & - I^\text{def} - \alpha_iO^\text{def}\\
& - (\text{COO}_{i,j} + \overline{\text{CSR}_{i,j}}) R^\text{def}_{-} + \text{CSR}_{i,j}R^\text{def}_{+}
\end{split}
\end{equation}
and
\begin{equation}
\begin{split}
\text{U}_{i,j}^\text{Bob} = & - I^\text{att} - \beta_jO^\text{att}\\
& - (\text{COO}_{i,j} + \overline{\text{ASR}_{i,j}}) R^\text{att}_{-} + \text{ASR}_{i,j}R^\text{att}_{+}.
\end{split}
\end{equation}

We consider the scenario in which the two players adopt mixed strategies, which indicates that the optimal responses for \emph{Alice} and \emph{Bob} correspond to a probability distribution over strategic space. As mentioned above, we reduce the number of pure strategies for both players to $2$. To this end, we rewrite $\mathcal{S}_A = \{\alpha_1, \alpha_2\}$ and $\mathcal{S}_B = \{\beta_1, \beta_2\}$. The action taken by \emph{Alice} can therefore be expressed as $(\text{Pr}\{\alpha_1\}, \text{Pr}\{\alpha_2\})$, where $\text{Pr}\{\alpha_1\}$ is the probability of choosing $\alpha_1$ as the defending strategy and $\text{Pr}\{\alpha_1\} + \text{Pr}\{\alpha_2\} = 1$. Similarly, the action taken by \emph{Bob} can be denoted by $(\text{Pr}\{\beta_1\}, \text{Pr}\{\beta_2\})$, where $\text{Pr}\{\beta_1\} + \text{Pr}\{\beta_2\} = 1$. Thus, the payoff matrix can be described as follows.
\[
\renewcommand{\arraystretch}{1.2}
\begin{array}{c|cc}
\text{Mixed strategy} & \text{Pr}\{\beta_1\} & \text{Pr}\{\beta_2\} \\
\hline
\text{Pr}\{\alpha_1\} & (\text{U}_{1,1}^{\text{Alice}},\text{U}_{1,1}^{\text{Bob}}) & (\text{U}_{1,2}^{\text{Alice}},\text{U}_{1,2}^{\text{Bob}}) \\
\text{Pr}\{\alpha_2\} & (\text{U}_{2,1}^{\text{Alice}},\text{U}_{2,1}^{\text{Bob}}) & (\text{U}_{2,2}^{\text{Alice}},\text{U}_{2,2}^{\text{Bob}}) \\
\end{array}
\]

In order to find the Nash equilibrium, the action taken by the present player should ensure that the expectation of the payoff of the adversary is unbiased. Mathematically, we expect that
\begin{equation}
\left\{\begin{matrix}
\mathbb{E}_{\text{Pr}\{\mathcal{S}_A\}}\left[\text{U}_{\star,1}^\text{Bob}\right] = \mathbb{E}_{\text{Pr}\{\mathcal{S}_A\}}\left[\text{U}_{\star,2}^\text{Bob}\right] \\
\mathbb{E}_{\text{Pr}\{\mathcal{S}_B\}}\left[\text{U}_{1,\star}^\text{Alice}\right] = \mathbb{E}_{\text{Pr}\{\mathcal{S}_B\}}\left[\text{U}_{2,\star}^\text{Alice}\right]
\end{matrix}\right., 
\end{equation}
which can be further detailed as:
\begin{equation}
\begin{split}
\text{Pr}\{\alpha_1\}\text{U}_{1,1}^\text{Bob} + \text{Pr}\{\alpha_2\}&\text{U}_{2,1}^\text{Bob} = \\
& \text{Pr}\{\alpha_1\}\text{U}_{1,2}^\text{Bob} + \text{Pr}\{\alpha_2\}\text{U}_{2,2}^\text{Bob}
\end{split}
\end{equation}
and
\begin{equation}
\begin{split}
\text{Pr}\{\beta_1\}\text{U}_{1,1}^\text{Alice} + \text{Pr}&\{\beta_2\}\text{U}_{1,2}^\text{Alice} = \\
& \text{Pr}\{\beta_1\}\text{U}_{2,1}^\text{Alice} + \text{Pr}\{\beta_2\}\text{U}_{2,2}^\text{Alice}.
\end{split}
\end{equation}

Obviously, we can find that
\begin{equation}
\text{Pr}\{\alpha_1\} = \frac{- \Delta \text{U}_{2,\star}^\text{Bob}}{\Delta \text{U}_{\star,1}^\text{Bob} - \Delta \text{U}_{\star,2}^\text{Bob}}
\end{equation}
and
\begin{equation}
\text{Pr}\{\beta_1\} = \frac{- \Delta \text{U}_{\star,2}^\text{Alice}}{\Delta \text{U}_{1,\star}^\text{Alice} - \Delta \text{U}_{2,\star}^\text{Alice}},
\end{equation}
where
\begin{equation}
\left\{\begin{matrix}
\begin{split}
\Delta \text{U}_{2,\star}^\text{Bob} & = \text{U}_{2,1}^\text{Bob} - \text{U}_{2,2}^\text{Bob}\\
\Delta \text{U}_{\star,1}^\text{Bob} & = \text{U}_{1,1}^\text{Bob} - \text{U}_{2,1}^\text{Bob}\\
\Delta \text{U}_{\star,2}^\text{Bob} & = \text{U}_{1,2}^\text{Bob} - \text{U}_{2,2}^\text{Bob}\\
\Delta \text{U}_{\star,2}^\text{Alice} & = \text{U}_{1,2}^\text{Alice} - \text{U}_{2,2}^\text{Alice}\\
\Delta \text{U}_{1,\star}^\text{Alice} & = \text{U}_{1,1}^\text{Alice} - \text{U}_{1,2}^\text{Alice}\\
\Delta \text{U}_{2,\star}^\text{Alice} & = \text{U}_{2,1}^\text{Alice} - \text{U}_{2,2}^\text{Alice}
\end{split}
\end{matrix}\right..
\end{equation}

By applying Eq. (16, 17) to Eq. (23), we further have
\begin{equation}
\begin{split}
\Delta \text{U}_{2,\star}^\text{Bob} & = \text{U}_{2,1}^\text{Bob} - \text{U}_{2,2}^\text{Bob}\\
& = (\beta_2 - \beta_1) O^\text{att} + (r_{2,2} - r_{2,1}) R_{+}^\text{att}\\
&~~~~~~~~~~~ - \left[k(\beta_1 - \beta_2) - (r_{2,2} - r_{2,1})\right] R_{-}^\text{att}
\end{split}
\end{equation}
\begin{equation}
\begin{split}
\Delta \text{U}_{\star,1}^\text{Bob} & = \text{U}_{1,1}^\text{Bob} - \text{U}_{2,1}^\text{Bob}\\
& = (r_{2,1} - r_{1,1})R_{+}^\text{att}\\
&~~~~~~~~~~~ - \left[ k(\alpha_1 - \alpha_2) - (r_{2,1} - r_{1,1}) \right] R_{-}^\text{att}
\end{split}
\end{equation}
\begin{equation}
\begin{split}
\Delta \text{U}_{\star,2}^\text{Bob} & = \text{U}_{1,2}^\text{Bob} - \text{U}_{2,2}^\text{Bob}\\
& = (r_{2,2} - r_{1,2})R_{+}^\text{att}\\
&~~~~~~~~~~~ - \left[ k(\alpha_1 - \alpha_2) - (r_{2,2} - r_{1,2}) \right] R_{-}^\text{att}
\end{split}
\end{equation}
\begin{equation}
\begin{split}
& \Delta \text{U}_{\star,2}^\text{Alice} = \text{U}_{1,2}^\text{Alice} - \text{U}_{2,2}^\text{Alice}\\
& = (\alpha_2 - \alpha_1)O^\text{def} + k(\alpha_2 - \alpha_1)R_{-}^\text{def} + \\
&~\left[ \lambda(1-\beta_2)(\alpha_2-\alpha_1) + \beta_2(r_{1,2} - r_{2,2}) \right](R_{+}^\text{def} + R_{-}^\text{def})
\end{split}
\end{equation}
\begin{equation}
\begin{split}
& \Delta \text{U}_{1,\star}^\text{Alice} = \text{U}_{1,1}^\text{Alice} - \text{U}_{1,2}^\text{Alice}\\
& = k(\beta_2 - \beta_1)R_{-}^\text{def} + [ (\beta_2-\beta_1) \\
&~~~~~~~~~~~ (1-\lambda\alpha_1) + (\beta_1r_{1,1}-\beta_2r_{1,2}) ](R_{+}^\text{def} + R_{-}^\text{def})
\end{split}
\end{equation}
\begin{equation}
\begin{split}
& \Delta \text{U}_{2,\star}^\text{Alice} = \text{U}_{2,1}^\text{Alice} - \text{U}_{2,2}^\text{Alice}\\
& = k(\beta_2 - \beta_1)R_{-}^\text{def} + [ (\beta_2-\beta_1) \\
&~~~~~~~~~~~ (1-\lambda\alpha_2) + (\beta_1r_{2,1}-\beta_2r_{2,2}) ](R_{+}^\text{def} + R_{-}^\text{def})
\end{split}
\end{equation}
from which we can find that the best responses for both players are independent of the initial costs $I^\text{def}$ and $I^\text{att}$. It can be easily determined that
\begin{equation}
\Delta \text{U}_{\star,1}^\text{Bob} - \Delta \text{U}_{\star,2}^\text{Bob} = (\Delta_{r_{\star, 2}} - \Delta_{r_{\star, 1}})(R_{+}^\text{att} + R_{-}^\text{att})
\end{equation}
and
\begin{equation}
\begin{split}
\Delta \text{U}_{1,\star}^\text{Alice} - \Delta \text{U}_{2,\star}^\text{Alice}  = (\lambda\Delta_\alpha&\Delta_\beta + \beta_1\Delta_{r_{\star,1}} \\
& - \beta_2\Delta_{r_{\star,2}})(R_{+}^\text{def} + R_{-}^\text{def}),
\end{split}
\end{equation}
where
\begin{equation}
\left\{\begin{matrix}
\begin{split}
\Delta_{r_{\star, 1}} & = r_{1,1} - r_{2,1}\\
\Delta_{r_{\star, 2}} & = r_{1,2} - r_{2,2}\\
\Delta_\alpha & = \alpha_1 - \alpha_2\\
\Delta_\beta & = \beta_1 - \beta_2
\end{split}
\end{matrix}\right..
\end{equation}
Similarly, we have
\begin{equation}
\Delta \text{U}_{2,\star}^\text{Bob} = - \Delta_{\beta}O^\text{att} - k\Delta_{\beta}R_{-}^\text{att} - \Delta_{r_{2,\star}}(R_{+}^\text{att}+R_{-}^\text{att})
\end{equation}
and
\begin{equation}
\begin{split}
\Delta \text{U}_{\star,2}^\text{Alice} = -\Delta_\alpha O^\text{def} & - k \Delta_\alpha R_{-}^\text{def} + [ -\lambda\Delta_\alpha\\
&  (1 - \beta_2) + \beta_2 \Delta_{r_{\star, 2}}] (R_{+}^\text{def}+R_{-}^\text{def}),
\end{split}
\end{equation}
where $\Delta_{r_{2, \star}} = r_{2,1} - r_{2,2}$. Consequently, we have
\begin{equation}
\text{Pr}\{\alpha_1\} = \frac{\Delta_{\beta}O^\text{att} + k\Delta_{\beta}R_{-}^\text{att} + \Delta_{r_{2,\star}}(R_{+}^\text{att}+R_{-}^\text{att})}{(\Delta_{r_{\star, 2}} - \Delta_{r_{\star, 1}})(R_{+}^\text{att} + R_{-}^\text{att})}
\end{equation}
and
\begin{equation}
\text{Pr}\{\beta_1\} = \frac{\Delta_\alpha O^\text{def} + k \Delta_\alpha R_{-}^\text{def} + \varrho(R_{+}^\text{def}+R_{-}^\text{def})}{\rho(R_{+}^\text{def} + R_{-}^\text{def})},
\end{equation}
where 
\begin{equation}
\varrho = \lambda\Delta_\alpha (1 - \beta_2)  - \beta_2 \Delta_{r_{\star, 2}}
\end{equation}
and
\begin{equation}
\rho = \lambda\Delta_\alpha\Delta_\beta + \beta_1\Delta_{r_{\star,1}} - \beta_2\Delta_{r_{\star,2}}.
\end{equation}
To facilitate the derivation, we further assume that $O^\text{att} = kR_{+}^\text{att}$ and $O^\text{def} = kR_{+}^\text{def}$, which can be reasonable since the ongoing defense or attack action determines the reward. Therefore,
\begin{equation}
\text{Pr}\{\alpha_1\} = \frac{k\Delta_{\beta} + \Delta_{r_{2,\star}}}{\Delta_{r_{\star, 2}} - \Delta_{r_{\star, 1}}}
\end{equation}
and
\begin{equation}
\text{Pr}\{\beta_1\} = \frac{k \Delta_\alpha + \varrho}{\rho}.
\end{equation}

Digital watermarking games often emphasize the conditions under which the defender adopts the specific strategies, while a comprehensive framework would ideally include an analysis of the attacker’s Nash equilibrium. However, in this study, the expression for $\text{Pr}\{\beta_1\}$, corresponding to the optimal response of \emph{Bob}, involves many parameters, making the direct analysis or derivation of equilibrium points challenging. In contrast, in Eq. (39), $\text{Pr}\{\alpha_1\}$, can be analyzed in a relatively easier way. 

\begin{figure}[!t]
\centering
\includegraphics[width=\linewidth]{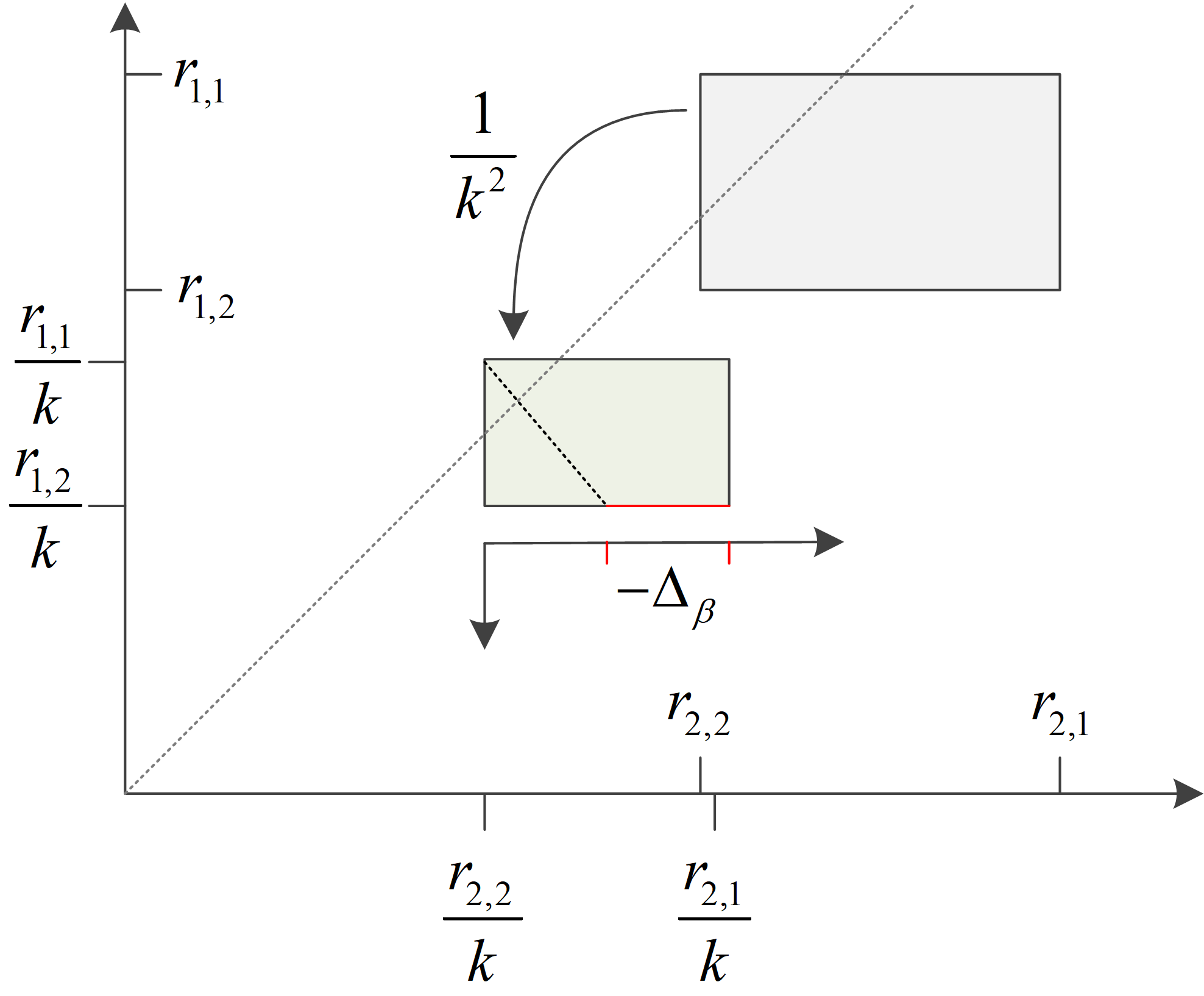}
\caption{The domain when the optimal response of \emph{Alice} is a mixed strategy.}
\label{fig:1}
\end{figure}

In Eq. (39), generally, we have $\Delta_\beta < 0$, $\Delta_{r_{2,\star}} > 0$, $\Delta_{r_{\star, 2}} < 0$, $\Delta_{r_{\star, 1}} < 0$, and $k > 0$. It is required that
\begin{equation}
\text{Pr}\{\alpha_1\} = \frac{k\Delta_{\beta} + \Delta_{r_{2,\star}}}{\Delta_{r_{\star, 2}} - \Delta_{r_{\star, 1}}} = \frac{k\Delta_{\beta} + \Delta_{r_{2,\star}}}{\Delta_{r_{2, \star}} - \Delta_{r_{1, \star}}} \in (0, 1).
\end{equation}
If $\Delta_{r_{2, \star}} - \Delta_{r_{1, \star}} > 0$, we have
\begin{equation}
- \Delta_{r_{2, \star}} < k\Delta_{\beta} < - \Delta_{r_{1, \star}},
\end{equation}
where $\Delta_{r_{1, \star}} = r_{1,1} - r_{1,2}$. Otherwise, we have
\begin{equation}
- \Delta_{r_{1, \star}} < k\Delta_{\beta} < - \Delta_{r_{2, \star}}
\end{equation}
for $\Delta_{r_{2, \star}} - \Delta_{r_{1, \star}} < 0$. So, a Nash equilibrium may occur if
\begin{equation}
\Delta_{\beta} \in \left (-\frac{\text{max}\{\Delta_{r_{1, \star}}, \Delta_{r_{2, \star}}\}}{k}, -\frac{\text{min}\{\Delta_{r_{1, \star}}, \Delta_{r_{2, \star}}\}}{k}\right ).
\end{equation}

Ignoring the constant coefficient $k$, the optimal response of \emph{Alice} is determined by the difference in attack intensity and the robustness differences of the two watermarking strategies. Fig. 1 illustrates the conditions under which the optimal response of \emph{Alice} is a mixed strategy when $\Delta_{r_{2, \star}} - \Delta_{r_{1, \star}} > 0$. It can be seen that the scaled robustness-differential rectangle, namely, the light green one, determines the range of the difference of the intensity between the two attacks to conduct the mixed action. There should be constraints for \emph{Bob}, which will reduce the parameter space and need to be investigated in the future. 

In summary, the proposed game reveals that the defender's strategy is subject to the difference between the robustness of different marked models and the difference between different attack strengths. This insight has motivated us to develop novel DNN model watermarking schemes to enhance the robustness against possible model attacks in the follow-up research.

\section{Conclusion}
In this paper, we investigate trigger-based model watermarking and propose a partial cooperation game framework. While conventional studies often only focus on either cooperative or non-cooperative games, we argue that there are both cooperative benefits and competitive benefits for the defender and the attacker. To model the game, we take into account economic costs and benefits for the definition of the payoff function. In the model, both players expect to maintain the performance of the model on the original task. However, they are competing with each other on watermark detection. The results indicate that the optimal response of the defender is subject to the robustness difference between different watermarked models and the strength difference between different attacks, highlighting the importance of enhancing the robustness of the watermarked model against real-world attacks during system design. 

Future work will focus on the impact of trigger set selection on DNN model performance in real-world scenarios. Practical implementations will be investigated to validate and extend the proposed framework, ensuring its robustness and applicability in diverse contexts. Moreover, generative model watermarking games will be studied to enrich the watermarking theory.

\section*{Acknowledgment}
This article was partly supported by the Science and Technology Commission of Shanghai Municipality (STCSM) under Grant Number 24ZR1424000, the Basic Research Program for Natural Science of Guizhou Province, and the National Natural Science Foundation of China under Grant Number U23B2023.


\end{document}